\documentclass[%
reprint,
superscriptaddress,
amsmath,amssymb,
aps,
longbibliography,
]{revtex4-1}

\usepackage[]{graphicx}
\usepackage{dcolumn}
\usepackage{bm}
\usepackage{verbatim}
\usepackage{mathrsfs}
\usepackage{hyperref}

\usepackage{color}

\begin{document}

    
\title{Crystalline topological Dirac semimetal phase in rutile structure $\beta'$-PtO$_2$}
    
\author{Rokyeon Kim}
\email{rrykim@gmail.com}
\affiliation{Department of Physics and Astronomy, Seoul National University, Seoul 08826, Korea}
\affiliation{Center for Correlated Electron Systems, Institute for Basic Science (IBS), Seoul 08826, Korea}

\author{Bohm-Jung Yang}
\affiliation{Department of Physics and Astronomy, Seoul National University, Seoul 08826, Korea}
\affiliation{Center for Correlated Electron Systems, Institute for Basic Science (IBS), Seoul 08826, Korea}
\affiliation{Center for Theoretical Physics (CTP), Seoul National University, Seoul 08826, Korea}

\author{Choong H. Kim}
\email[]{chkim82@snu.ac.kr}
\affiliation{Department of Physics and Astronomy, Seoul National University, Seoul 08826, Korea}
\affiliation{Center for Correlated Electron Systems, Institute for Basic Science (IBS), Seoul 08826, Korea}
    
\date{\today}

\begin{abstract}
Based on first-principles calculations and symmetry analysis, we propose that a transition metal rutile oxide, in particular $\beta'$-PtO$_2$, can host a three-dimensional topological Dirac semimetal phase.
We find that $\beta'$-PtO$_2$ possesses an inner nodal chain structure when spin-orbit coupling is neglected.
Incorporating spin-orbit coupling gaps the nodal chain, while preserving a single pair of three-dimensional Dirac points protected by a screw rotation symmetry.
These Dirac points are created by a band inversion of two $d$ bands, which is a realization of a Dirac semimetal phase in a correlated electron system.
Moreover, a mirror plane in the momentum space carries a nontrivial mirror Chern number $n_M = -2$, which distinguishes $\beta'$-PtO$_2$ from the Dirac semimetals known so far, such as Na$_3$Bi and Cd$_3$As$_2$.
If we apply a perturbation that breaks the rotation symmetry and preserves the mirror symmetry, the Dirac points are gapped and the system becomes a topological crystalline insulator.
\end{abstract}


\maketitle
    

\section{Introduction}

Three-dimensional (3D) Dirac semimetals (DSMs) are materials that are characterized by fourfold degenerate nodal points or lines.
The low-energy excitations of DSMs describe Dirac fermions, and this direct correspondence to the elementary particle enables us to explore high-energy physics in condensed matter systems.
Especially, DSMs can carry nontrivial topological numbers, which lead to the observations of intriguing effects such as the surface Fermi arcs~\cite{Xu2015}, giant magnetoresistance~\cite{Liang:2015ev}, and quantum oscillations~\cite{Moll2016}.

Because of these unique properties, comprehensive theoretical studies were carried out on the existence and protection of Dirac points in materials.
For the systems without spin-orbit coupling (SOC), inversion ($P$) and time-reversal ($T$) symmetries protect nodal lines owing to the quantized Berry phases~\cite{Kim2015, Yu2015}.
When SOC is included, however, the nodal lines are gapped, and we require additional crystalline symmetries to protect Dirac points.
The protection of Dirac points in the presence SOC can be classified into two cases, namely, the band inversion and the symmetry enforced mechanisms.
They can be explained in a unified way by inspecting the rotation symmetries of crystals~\cite{Yang:2014ia}.
When a crystal has a rotation symmetry (including a screw rotation), a pair of Dirac points can be protected on the rotation axis via the band inversion mechanism~\cite{Wang:2012ds, Wang:2013is},
whereas when a crystal has a screw rotation symmetry, a single Dirac point is enforced at the Brillouin zone boundary.~\cite{Young:2012kz, Steinberg:2014ek}.

For the first case, when we choose the $k_z$ axis as a rotation axis, the $k_z= \text{0 or } \pi$ plane can carry two-dimensional (2D) topological invariants~\cite{Yang:2014ia}.
For example, DSMs Na$_3$Bi~\cite{Wang:2012ds} and Cd$_3$As$_2$~\cite{Wang:2013is}, which have been verified in experiments~\cite{Liu:2014bf,Neupane:2014kc,Liu:2014hr,Jeon2014}, are shown to have nontrivial $\mathbb{Z}_2$ invariants.
While the theory also predicts a mirror Chern number $n_M = \left| 2 \right|$ (and $ \left| 3 \right| $)~\cite{Yang:2014ia}, until recently there have been no studies on the DSMs with a mirror Chern number $n_M = \left| 2 \right|$.
A DSM carrying a nontrivial mirror Chern number, which we call a \emph{crystalline} topological DSM, is topologically distinct from the conventional DSM possessing a nontrivial $\mathbb{Z}_2$ number protected by $T$ symmetry.
To our knowledge, only VAl$_3$ was proposed to be a type-II DSM that has a mirror Chern number $n_M = 2$~\cite{Chang:2017gp}.
In contrast to a type-I DSM, where the Fermi surface shrinks to isolated points, a type-II DSM features the Fermi surface composed of electron and hole pockets, which is due to a tilted Dirac cone~\cite{Soluyanov2015}.

In this paper, we show that the rutile phase PtO$_2$, $\beta^\prime$-PtO$_2$, is a type-I 3D topological DSM that possesses a nontrivial mirror Chern number $n_M = -2$.
We carried out the density functional theory (DFT) calculations and symmetry analysis, which reveal the protection mechanism of the Dirac points and the topological nature of the system.
Recently, a number of theoretical studies have been performed on the topological phases of the rutile oxides with transition metal ions.
For example, the Chern insulating state~\cite{Huang:2015ha} and the quantum spin Hall phase~\cite{Lado:2016bh} were predicted in the rutile-based heterostructures.
In addition, for the 3D bulks, IrO$_2$ was shown to has Dirac nodal lines on the zone boundary~\cite{Sun:2017dh},
and $\beta$-PbO$_2$ was proposed to be a 3D DSM~\cite{Wang:2017eq, Wang:2017jp}.

Compared with the DSMs studied so far, the DSM phase of $\beta'$-PtO$_2$ shows distinct features.
Unlike $\beta$-PbO$_2$ and other DMSs, which usually involves $s$ and $p$ bands, the DSM phase of $\beta^\prime$-PtO$_2$ is realized in a correlated electron system by a band inversion of two $d$ bands.
Moreover, as the system carries a nontrivial mirror Chern number, 2D topological phases~\cite{Huang:2015ha,Lado:2016bh} as well as the topological crystalline insulator (TCI) phase can be achieved in $\beta^\prime$-PtO$_2$ by reducing the dimension and breaking the symmetries.
Therefore, the crystal structures based on $\beta^\prime$-PtO$_2$ may serve as a platform to study the correlation effects and the various topological phases in DSMs.
The possible influence of electronic correlations in DSMs includes the correlation-induced semimetal-insulator transitions~\cite{Sekine2014, Roy2017a,Han2018} and unusual quantum critical transports~\cite{Goswami2011,Hosur2012,Li2017}.
Lastly, we point out that the mirror symmetry $M_z \colon z \to -z$ in the rutile structure is overlooked in Refs.~\cite{Lado:2016bh, Wang:2017eq, Wang:2017jp}, and they missed the correct description of 2D topological invariants, which we present in this study.

\section{Method and Crystal structure}

To investigate the electronic structures of the transition metal rutile oxides, we have carried out the DFT calculations using the Vienna \textit{ab initio} simulation code~\cite{Kresse1996}.
The projector-augmented-wave method~\cite{Blochl1994} and the exchange correlation functional of the generalized gradient approximation (GGA) in the Perdew, Burke, and Ernzerhof~\cite{Perdew1996a} scheme were used.
To improve the estimation of a band gap, we further employed the Heyd-Scuseria-Ernzerhof (HSE) hybrid functional calculations~\cite{Heyd2003}.
The self-consistent total energy was evaluated with the $8 \times 8 \times 12$ $k$-point mesh and the cutoff energy for the plane-wave basis set was 500 eV.

A transition metal rutile oxide, $M$O$_2$, has a tetragonal structure with space group $P4_2/mnm$ (No. 136).
This compound can contain group VIII transition metals, such as Rh, Pd, Ir, and Pt.
Especially, platinum dioxides can be crystallized in the rutile phase $\beta^\prime$-PtO$_2$~\cite{Fernandez:1984ga} along with allotropes $\beta$-PtO$_2$ (space group $Pnnm$)~\cite{Shannon1968} and $\alpha$-PtO$_2$ (space group $P\bar{3}m1$)~\cite{Mansour1984}.
Though the calculated energy of $\beta$-PtO$_2$ is the lowest of the three, the energy differences between the structures are relatively small~\cite{Nomiyama:2011ex}.
Experimentally, when high O$_2$ pressure in the range of 40 to 60 kbar is used, the rutile phase $\beta^\prime$-PtO$_2$ can be stabilized among the three structures~\cite{Fernandez:1984ga}.

\begin{figure}[]
\centering
\includegraphics[width=\columnwidth]{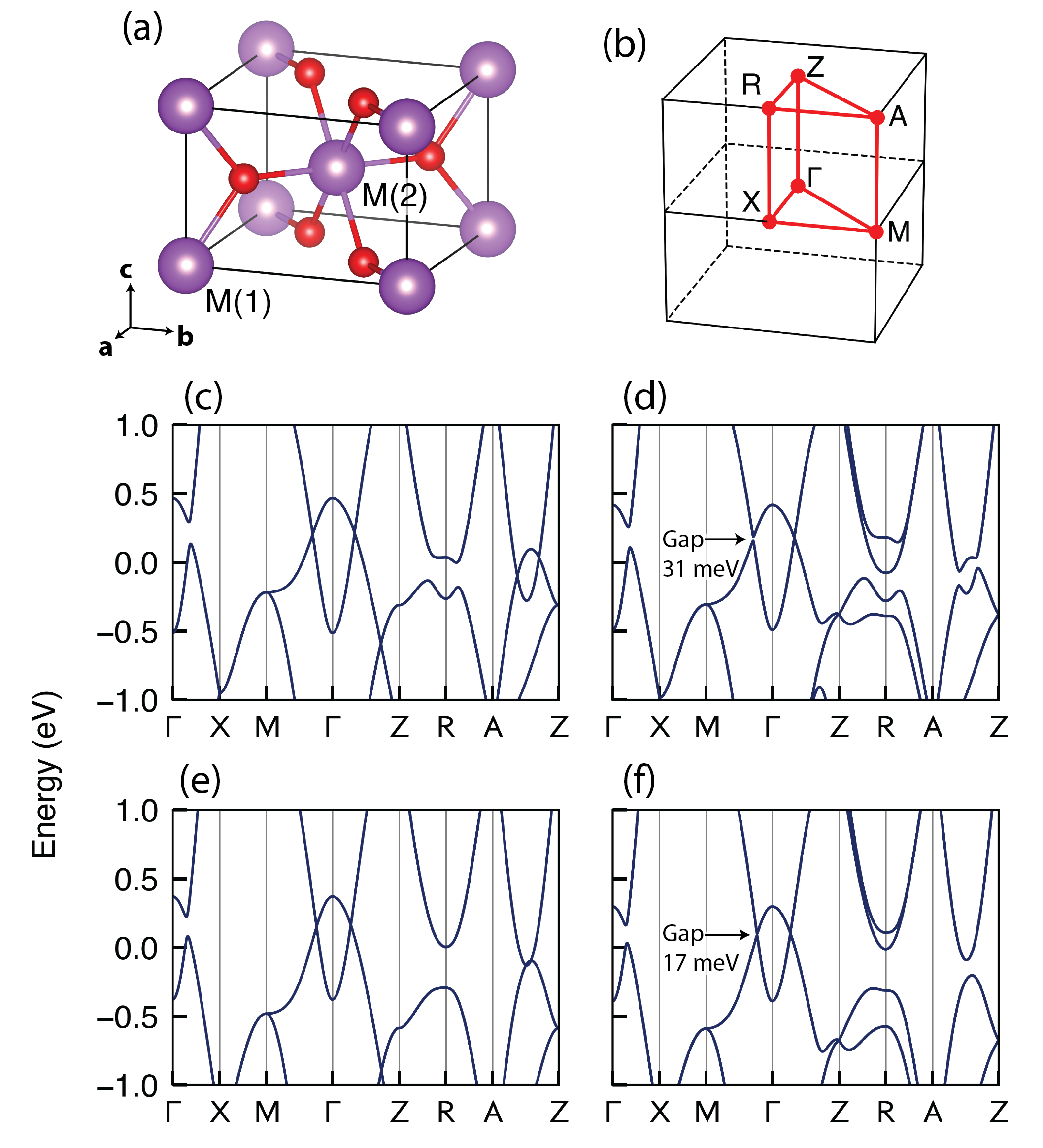}
\caption{(a) Crystal structure of the rutile oxides. 
(b) Tetragonal Brillouin zone.
(c) GGA  and (d) GGA + SOC band structures of $\beta'$-PtO$_2$.
(e) GGA  and (f) GGA + SOC band structures of $\beta'$-PtO$_2$ uniformly  compressed by 3~\%.}
\label{Fig_1}
\end{figure}

In the unitcell of the rutile structure, two cations, $M(1)$ and $M(2)$, are displaced by $\bm{\tau}=(\frac{1}{2},\frac{1}{2},\frac{1}{2})$ (in units of ($a$,$a$,$c$)) from one another, and each cation is surrounded by six oxygen atoms as shown in Fig. \ref{Fig_1}(a).
The space group includes five generators: $E$,  $C_{2z}$, $\widetilde{C}_{4z}=\{ C_{4z}|\bm{\tau} \}$, $\widetilde{C}_{2y}=\{ C_{2y}|\bm{\tau} \}$, and $P$,
where $E$ is the identity, $C_{2z}$ is a twofold rotation about the $z$ axis, 
and $\widetilde{C}_{4z}$ ($\widetilde{C}_{2y}$) is a fourfold (twofold) rotation about the  $z$ axis ($y$ axis), $C_{4z}$ (${C}_{2y}$), followed by a translation by $\bm{\tau}$.
Among the 16 symmetry operations constructed from the generators, the following symmetries are important in this study:
\begin{equation*}
P ,\; \widetilde{C}_{4z} ,\; M_z=M_{\left[001\right]} ,\; M_{\left[110\right]} ,\; M_{\left[1\bar{1}0\right]},
\end{equation*}
where $M_\mathbf{n}$ is a mirror reflection about the plane defined by a normal vector $\mathbf{n}$.
Additionally, as we focus on non-magnetic systems, we include $T$ symmetry.

\section{Results}

Figures~\ref{Fig_1}(c) and \ref{Fig_1}(d) show the GGA and GGA + SOC band structures, respectively, of $\beta^\prime$-PtO$_2$ with the experimental lattice constants $a=4.485$~{\AA}, $c=3.130$~{\AA}~\cite{Fernandez:1984ga}.
Without SOC we have nodal points near the Fermi level on the M{\textendash}$\Gamma$, $\Gamma${\textendash}Z, and A{\textendash}Z lines [Fig.~\ref{Fig_1}(c)].
As the Pt ion in the rutile structure has a $5d^6$ configuration and octahedral-like coordination, the nodal points are created by a band inversion of $t_{2g}$ and $e_g$ bands.  
When we include SOC, however, the nodal points are gapped except the ones on the $\Gamma${\textendash}Z line [Fig.~\ref{Fig_1}(d)].
In other words, in the presence of SOC, we observe the 3D DSM phase of $\beta^\prime$-PtO$_2$, which has a band gap locally at every $\mathbf{k}$ point except the Dirac points on the $\Gamma${\textendash}Z line.
We note, however, that the Dirac points are slightly above the Fermi level because of the bands on the $k_z = \pi$ plane. 
Although the system can be utilized as a DSM in its present condition, we can additionally tune the band structure to move the Dirac points close to the Fermi level.
We find that the bands on the $k_z = \pi$ plane can be removed by applying a uniform compressive strain.
For example, when we apply a uniform compressive strain of 3~{\%}, the bands on the $k_z = \pi$ plane are pushed away and the Dirac points get closer to the Fermi level as we can see in the GGA [Figs.~\ref{Fig_1}(e)] and GGA + SOC [Figs.~\ref{Fig_1}(f)] band structures.
Having revealed that $\beta^\prime$-PtO$_2$ hosts a DSM phase, we investigate the protection mechanism of the Dirac points and the topological properties of the system in the subsequent sections.

\subsection{Dirac Nodal Lines without SOC}
We begin by reviewing the general properties of a band structure with symmetries.
The presence of both $P$ and $T$ symmetries ensures that the band structure is symmetric in $\mathbf{k}$ and $-\mathbf{k}$ points, and is doubly degenerate at every $\mathbf{k}$ point.
In general, when we consider a symmetry operation $\widetilde{g}=\{g|\mathbf{t}\}$ of a system,
the invariance of the system under $\widetilde{g}$ requires that the Bloch Hamiltonian satisfies $ \widetilde{g} H(\mathbf{k}) \widetilde{g}^{-1}= H(R_{g}\mathbf{k})$, where $R_{g}$ is a rotation matrix representing $g$.
As a result, for the $\mathbf{k}$ point that is invariant under $R_{g}$, i.e., $R_{g} \mathbf{k}=\mathbf{k}$, $\tilde{g}$  and $H(\mathbf{k})$ commute, and a Bloch state can be labeled by a $\widetilde{g}$ eigenvalue.

\begin{figure}[]
\centering
\includegraphics[width=\columnwidth]{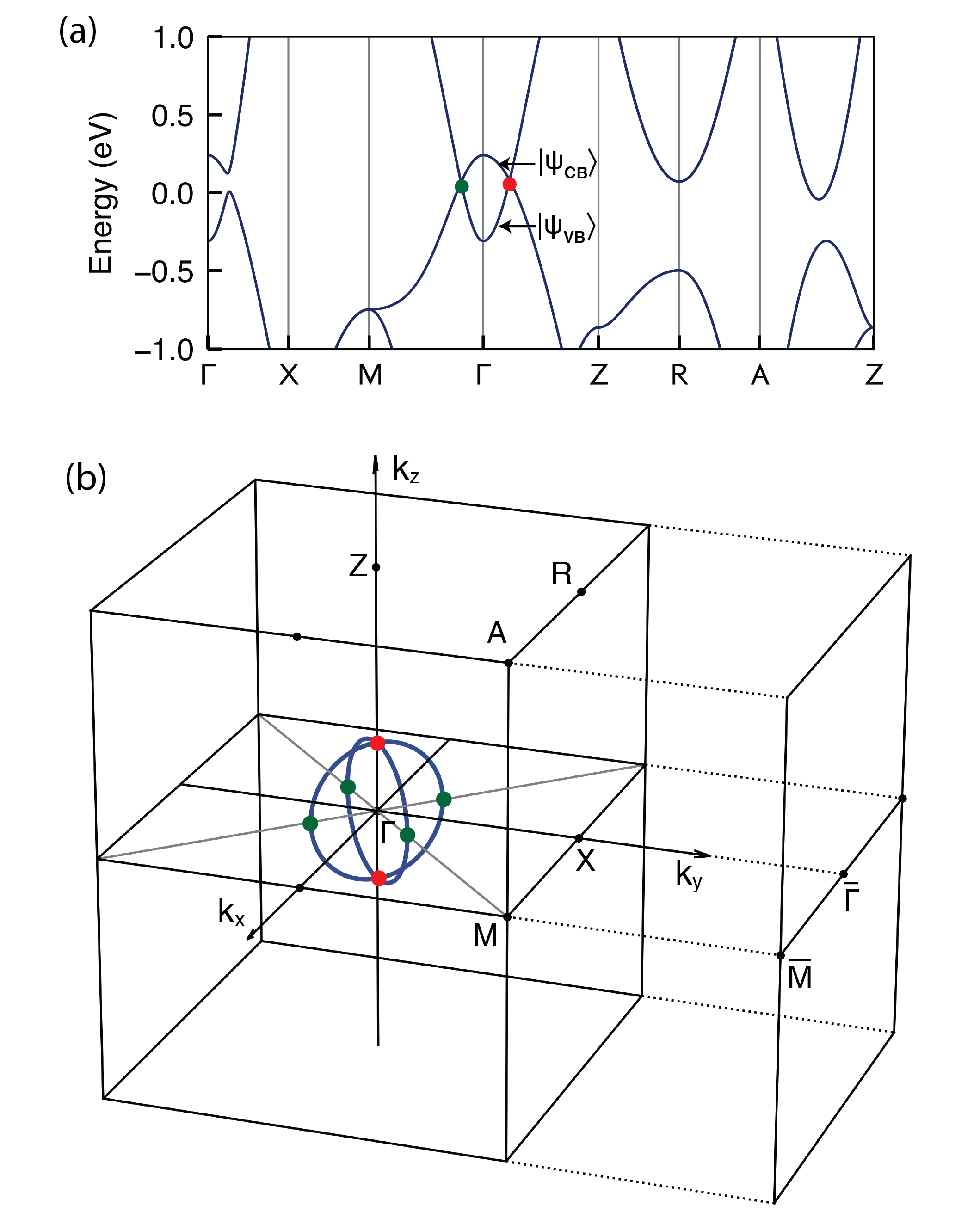}
\caption{(a) GGA band structure of $\beta^\prime$-PtO$_2$ uniformly compressed by 5~\%.
The valence band ($\left| \psi_\mathrm{VB} \right>$) is in contact with the conduction band ($\left| \psi_\mathrm{CB} \right>$) at the Dirac points (green and red dots).
(b) Whole Dirac nodal lines structure without SOC.
Two ring-shaped nodal lines on the (110) and (1$\bar{1}$0) planes join at the two points on the $k_z$ axis (red dots).
The bulk Brillouin zone and the surface Brillouin zone projected on the (010) surface are shown.}
\label{Fig_2}
\end{figure}

Figure \ref{Fig_2}(a) shows the band structure of $\beta^\prime$-PtO$_2$ in the absence of SOC ($T^2=1$).
Here, to focus on the band structure on the $k_z = 0$ plane and $k_z$ axis, we apply a uniform compressive strain of 5~{\%} to the system, but we emphasize that the system shows a DSM phase even without any strain.
In the band structure, we observe nodal points on the $\Gamma$\textendash{M} and $\Gamma$\textendash{Z} lines at ($k_0$,$k_0$,0) and (0,0,$k_d$), respectively, where $k_0=0.20$ $\pi$ and $k_d=0.23$ $\pi$ (momentums are in units of ($\frac{1}{a}$,$\frac{1}{a}$,$\frac{1}{c}$)).
In fact, the nodal points are parts of the nodal lines in the $(110)$ and $(1\bar{1}0)$ planes as shown shown in Fig. \ref{Fig_2}(b).
The whole nodal lines structure exhibits two ring-shaped nodal lines in the $(110)$ and $(1\bar{1}0)$ planes that are touching at $k_z= \pm k_d$.

To explain the nodal lines structure, we first consider the screw rotation $\widetilde{C}_{4z}$.
Under the successive operations of $\widetilde{C}_{4z}$ on a coordinate $(x,y,z)$, 
we have $\widetilde{C}_{4z}^4 : (x,y,z) \rightarrow (x,y,z+2)$.
Thus we can write $\widetilde{C}_{4z}^{4} =\tau(0,0,2)$, where $\tau(\alpha,\beta,\gamma)$ ($\alpha,\beta,\gamma \in \mathbb{Z}$) is a translation operator.
Since the effect of the translation operator acting on a Bloch state is
$ \tau(\alpha,\beta,\gamma)|\psi(\mathbf{k})\rangle = e^{-i(\alpha k_x + \beta k_y + \gamma k_z)} |\psi(\mathbf{k})\rangle $,
we obtain $\widetilde{C}_{4z}^4=e^{-i2k_z}$.
As a result, for the Bloch states on the $k_z$ axis,  $\widetilde{C}_{4z}$ eigenvalues are give by $e^{-i\frac{k_z}{2}}e^{i\frac{\pi}{2}n}$ $(n=0,1,2,3)$.

For the band structure of $\beta^\prime$-PtO$_2$, two bands near the Fermi level, which we denote $|\psi_\mathrm{VB}\rangle$ and $|\psi_\mathrm{CB}\rangle$ (see Fig. \ref{Fig_2}(a)), are mostly composed of the $d_{xy}$ and $d_{x^2-y^2}$ orbitals, respectively. 
On the $k_z$ axis, these bands can be written in the orbital basis as
\begin{equation*}
\begin{split}
|\psi_\mathrm{VB}\rangle &= \sum_{\mathbf{R}}  e^{i k_z R_z } \left[ |d_{xy}\rangle_\mathrm{Pt(1)} + e^{i \frac{k_z}{2} } |d_{xy}\rangle_\mathrm{Pt(2)} \right]_\mathbf{R} , \\
|\psi_\mathrm{CB}\rangle &= \sum_{\mathbf{R}} e^{i k_z R_z } \left[ |d_{x^2-y^2}\rangle_\mathrm{Pt(1)} - e^{i \frac{k_z}{2}} |d_{x^2-y^2}\rangle_\mathrm{Pt(2)}  \right]_\mathbf{R}  ,
\end{split}
\end{equation*}
where $\mathbf{R}$ is a lattice vector and other orbitals are omitted for simplicity.
Because of the nonsymmorphic character of the $\widetilde{C}_{4z}$ symmetry,
$|\psi_\mathrm{VB} \rangle$ and $|\psi_\mathrm{CB} \rangle$ have $\widetilde{C}_{4z}$ eigenvalues $-e^{-i\frac{{k}_{z}}{2}}$ and $+e^{-i\frac{{k}_{z}}{2}}$, respectively.
As the two bands belong to the different eigenvalues, they can cross on the $k_z$ axis without a hybridization and can form a pair of Dirac points at $k_z = \pm k_d$.

Next, we examine the mirror operations $M_{[110]}$ and $ M_{[1\bar{1}0]}$.
Since the $(110)$ and $(1\bar{1}0)$ planes are invariant under $M_{[110]}$ and $ M_{[1\bar{1}0]}$, respectively, we can label the Bloch states on these planes with mirror eigenvalues.
For each mirror plane, $|\psi_\mathrm{VB}\rangle$ and $|\psi_\mathrm{CB}\rangle$ have mirror eigenvalues $+1$ and $-1$, respectively, therefore they can form a nodal line on the mirror plane.
The two nodal lines on the $(110)$ and $(1\bar{1}0)$ planes are in the identical shape because of the $\widetilde{C}_{4z}$ symmetry.
Moreover, as the mirror planes contain the $k_z$ axis, the nodal lines should pass through the points (0,0,$+k_{d}$) and (0,0,$-k_{d}$).
Consequently, without SOC, we have an inner nodal chain structure~\cite{Bzdusek2016,Chang2017} as shown in Fig. \ref{Fig_2}(b). 

\subsection{Protection of Dirac Points with SOC}
\begin{figure}[]
\centering
\includegraphics[width=\columnwidth]{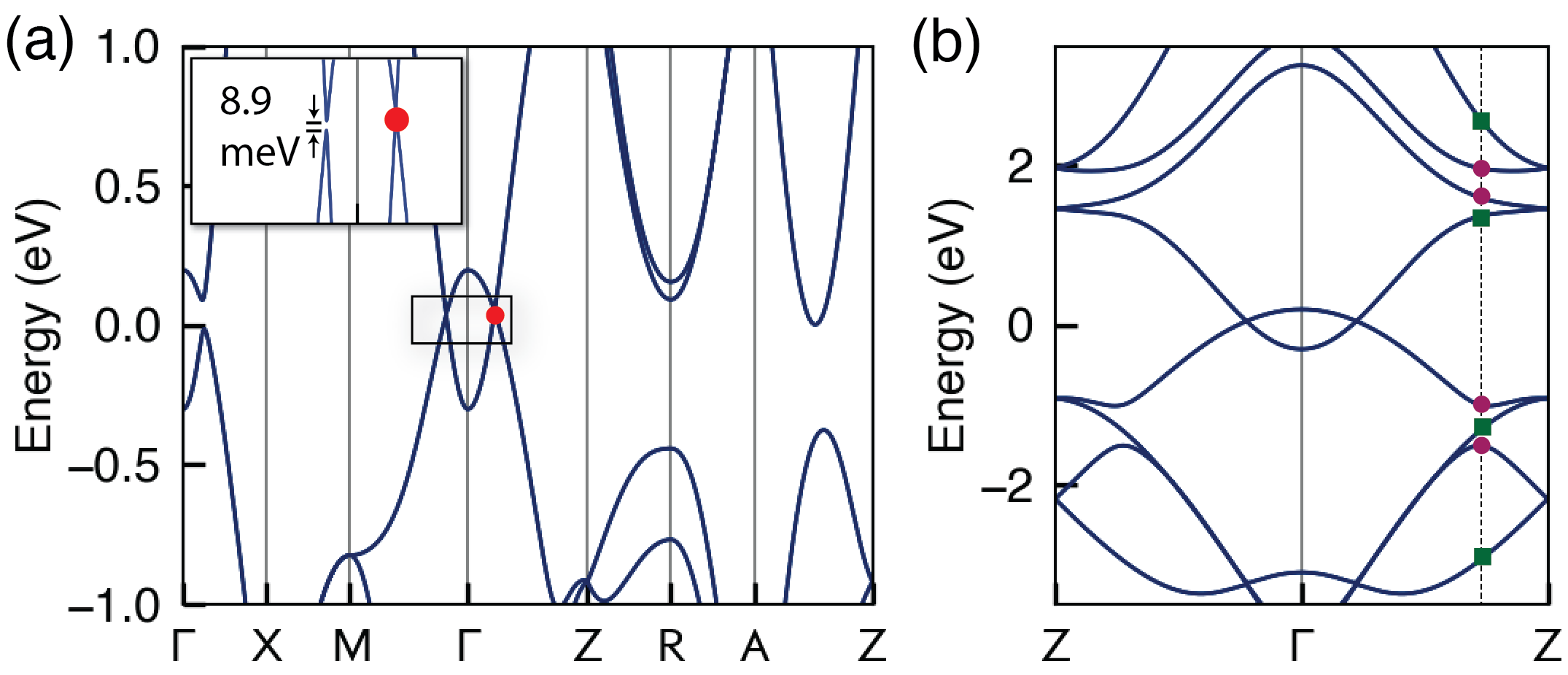}
\caption{(a) GGA + SOC  band structure of $\beta^\prime$-PtO$_2$ uniformly compressed by 5~\%.
The Dirac nodal lines are gapped except for the Dirac points on the $k_z$ axis (red dot).
(b) Bands on the $k_z$ axis are labeled by $\widetilde{C}_{4z}$ eigenvalues.
Round (purple) dots denote $\{\widetilde{c}_{4z}(0),\widetilde{c}_{4z}(3)\}$  and square (green) dots denote $\{\widetilde{c}_{4z}(1),\widetilde{c}_{4z}(2)\}$.}
\label{Fig_3}
\end{figure}

In the presence of SOC ($T^2=-1$), the nodal lines in the mirror planes are gapped as we can observe in the GGA + SOC band structure in Fig. \ref{Fig_3}(a).
We can explain the instability of the nodal lines by inspecting the mirror eigenvalues of the bands with SOC.
We first note that mirror eigenvalues take two imaginary values $\pm i$.
In general, on a plane that is invariant under a mirror operation $M_\mathbf{n}$, a Bloch state $|\psi(\mathbf{k})\rangle$ and its degenerate partner $PT|\psi(\mathbf{k})\rangle$ have different mirror eigenvalues because of the commutation relations $[P,T]=[M_\mathbf{n},T]=[P,M_\mathbf{n}]=0$.
Therefore, when two sets of degenerate bands approach on the mirror plane, there is an unavoidable hybridization between them and band gaps must open.

The Dirac points on the $k_z$ axis, however, are protected even with SOC (see Fig. \ref{Fig_3}(a)), which is due to the $\widetilde{C}_{4z}$ symmetry.
First, with SOC,  we have $\widetilde{C}_{4z}^{4}=e^{-i2k_z}(-1)$,
where the $-1$ is from the $2\pi$ rotation of a half-integer spin.
Therefore $\widetilde{C}_{4z}$ eigenvalues take
\begin{equation*}
\widetilde{c}_{4z} (n) = e^{-i\frac{k_z}{2}}e^{i\frac{\pi}{2}(n+\frac{1}{2})} \; (n=0,1,2,3) .
\end{equation*}
Next, because of the nonsymmorphic nature of $\widetilde{C}_{4z}$, we have the relation   $\widetilde{C}_{4z}P=\tau(1,1,1)P\widetilde{C}_{4z}$ in addition to the commutation relations $[T,P]=[T,\widetilde{C}_{4z}]=0$.
Then, for a Bloch state $|\psi(\mathbf{k})\rangle$ on the $k_z$ axis with a $ \widetilde{C}_{4z}$ eigenvalue $e^{-i\frac{k_z}{2}} e^{i\frac{\pi}{2}(n+\frac{1}{2})}$, we have
\begin{equation*}
\begin{split}
\widetilde{C}_{4z}PT|\psi(\mathbf{k})\rangle
& = T \tau(1,1,1) P \widetilde{C}_{4z} |\psi(\mathbf{k})\rangle \\
& = e^{-i\frac{k_z}{2}}e^{-i\frac{\pi}{2}(n+\frac{1}{2})}PT|\psi(\mathbf{k})\rangle \ .
\end{split}
\end{equation*}
As a result, we find that a degenerate doublet $\{ |\psi(\mathbf{k})\rangle, PT|\psi(\mathbf{k})\rangle \}$ of the Bloch Hamiltonian can have a pair of $\widetilde{C}_{4z}$ eigenvalues $\{\widetilde{c}_{4z}(0),\widetilde{c}_{4z}(3)\}$ or $\{\widetilde{c}_{4z}(1),\widetilde{c}_{4z}(2)\}$.

With $\widetilde{C}_{4z}$ eigenvalues of degenerate doublets, we can identify two classes of Dirac points on the $k_z$ axis.
First, if a doublet has $\widetilde{C}_{4z}$ eigenvalues $\{\widetilde{c}_{4z}(0),\widetilde{c}_{4z}(3) \}$,
it should touch with another doublet with $\{\widetilde{c}_{4z}(1),\widetilde{c}_{4z}(2) \}$ at the zone boundary $k_z= \pm \pi$ because of the connectivity conditions imposed on $\widetilde{c}_{4z}(n)$~\cite{Yang:2015er}.
Namely, two doublets whose four bands have all different $\widetilde{C}_{4z}$ eigenvalues stick together at the zone boundary.
To be more specific, at the $k_z= \pm \pi$ point, four degenerate states $\{ |\psi(\mathbf{k})\rangle, P|\psi(\mathbf{k})\rangle, T|\psi(\mathbf{k})\rangle, PT|\psi(\mathbf{k})\rangle \}$ can exhaust all $\widetilde{C}_{4z}$ eigenvalues because of the anticommutation relation $P \widetilde{C}_{4z} =  - \widetilde{C}_{4z} P $ at this point.
Thus for the systems with the $\widetilde{C}_{4z}$ symmetry, Dirac points at $(0,0,\pm \pi)$ is enforced and protected by the symmetry.
Secondly, Dirac points can be created by the band inversion mechanism that the sequence of bands is inverted in some regions of the momentum space.
When a band inversion occurs for two doublets that have different $\widetilde{C}_{4z}$ eigenvalues, there is no hybridization between them, and a pair Dirac points can be stabilized at general points on the $k_z$ axis.

For $\beta^\prime$-PtO$_2$, the Dirac points $(0,0,\pm k_d)$ fall into the second case.
The $\widetilde{C}_{4z}$ eigenvalues of $|\psi_\mathrm{VB} \rangle$ are $\{\widetilde{c}_{4z}(1),\widetilde{c}_{4z}(2)\}$ and those of $|\psi_\mathrm{CB}\rangle$ are $\{\widetilde{c}_{4z}(0),\widetilde{c}_{4z}(3)\}$, therefor the Dirac points can be protected even with SOC as shown in Fig.~\ref{Fig_3}(a).
Besides these Dirac points near the Fermi level, we have other Dirac points, both the first and second classes, in the whole energy range that are protected by $\widetilde{C}_{4z}$.
We label the bands on the $k_z$ axis with $\widetilde{C}_{4z}$ eigenvalues in Fig.~\ref{Fig_3}(b), which demonstrates the two classes of Dirac points created by the $\widetilde{C}_{4z}$ symmetry.
At the zone boundary, we can observe that all points are fourfold degenerate, which is due to the nonsymmorphic nature of $\widetilde{C}_{4z}$.
Additionally, on the general points on the $k_z$ axis, the band crossing is allowed when two doublets have different $\widetilde{C}_{4z}$ eigenvalues.
In this regard, $\beta^\prime$-PtO$_2$ is an intriguing system that we can simultaneously observe both classes of Dirac points. 

\subsection{Topological Properties}
\begin{figure}[]
\centering
\includegraphics[width=\columnwidth]{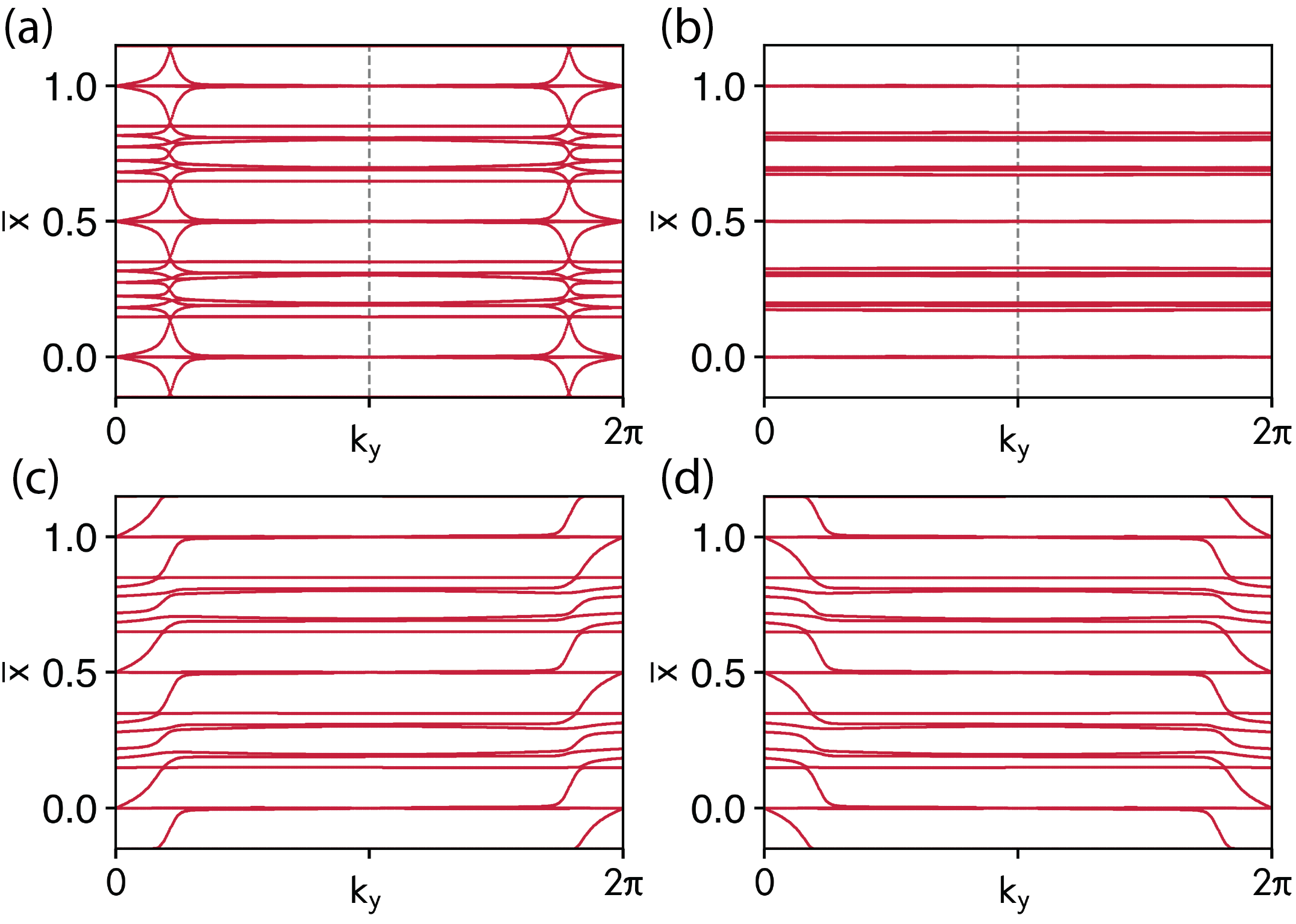}
\caption{Flow of the WCCs in the (a) $k_z = 0$ and (b) $k_z = \pi$ planes.
They exhibit even numbers of crossings in the half Brillouin zone, therefore are $\mathbb{Z}_2$ trivial.
For the $k_z = 0$ plane, the WCCs are calculated for each mirror eigensector (c) $+i$ and (d) $-i$. The Chern number of the $\pm i$ eigensector is $\mp 2$ indicated by the pumping of two electrons in one period.
The sign of the Chern number is determined from the propagation direction of the surface bands in Fig.~\ref{Fig_5}.
As a result the $k_z=0$ plane is described by a mirror Chern number $n_M = -2$.
}
\label{Fig_4}
\end{figure}

Having identified that $\beta'$-PtO$_2$ is a stable 3D DSM even with SOC, we now scrutinize topological properties of the system.
As the Dirac points are located at $k_z = \pm k_d$, the system has a full gap on the $k=0 \text{ and } \pi$ planes.
These planes can be considered as 2D systems with $T$ symmetry~\cite{Moore2007,Fu2007a}, additionally they are invariant under the $M_z$ operation.
Therefore we can define $\mathbb{Z}_2$ invariants ~\cite{Kane:2005gb} and mirror Chern numbers~\cite{Teo:2008bk} on the planes~\cite{Yang:2014ia}.
By calculating the flow of the Wannier charge centers (WCCs) on the $k_z=0 \text{ and } \pi $ planes, we find that both planes are trivial in terms of the $\mathbb{Z}_2$ invariant (see Figs.~\ref{Fig_4}(a) and \ref{Fig_4}(b)) \cite{Soluyanov2011, Yu2011}.
This result can also be deduced from the same parity of $|\psi_\mathrm{VB} \rangle$ and $|\psi_\mathrm{CB} \rangle$, which are mostly composed of the $d$ orbitals~\cite{Fu:2007eia}.
We discover, however, that the $k_z=0$ plane has a nontrivial mirror Chern number $n_M = -2$ (see Figs.~\ref{Fig_4}(c) and \ref{Fig_4}(d)), while the $k_z=\pi$ plane is trivial.
The mirror Chern number $n_M = -2$ of $\beta'$-PtO$_2$ distinguishes it from other DSMs such as Cd$_3$As$_2$ and Na$_3$Bi, which carry nontrivial $\mathbb{Z}_2$ invariants.
If we introduce a surface perpendicular to the mirror plane, multiple surface states corresponding to the mirror Chern number emerge at the boundary.
This surface states is different from the Fermi arcs in Weyl semimetals because it is solely determined by the 2D topological invariant in the $k_z=0$ plane.

\begin{figure}[]
\centering
\includegraphics[width=\columnwidth]{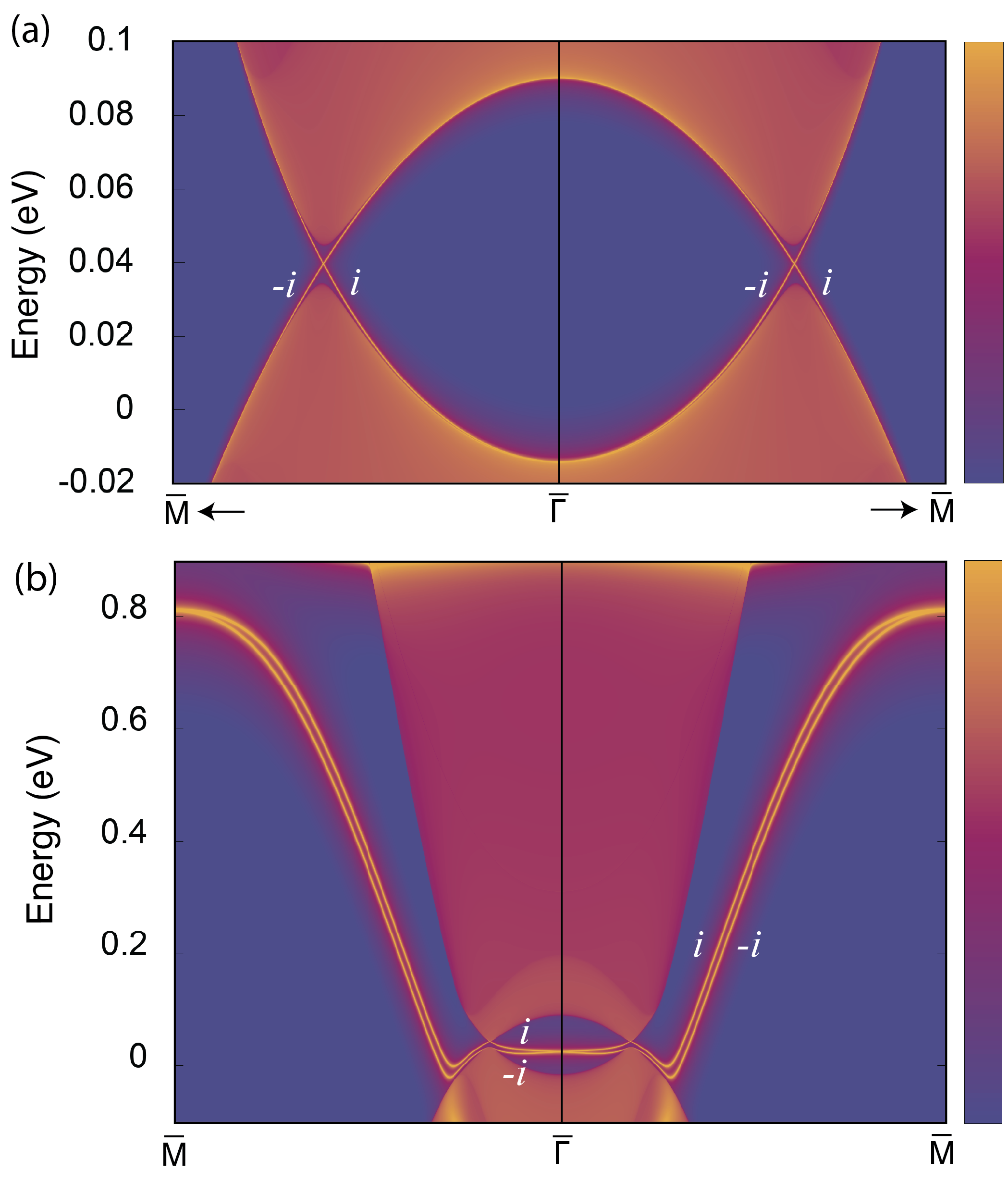}
\caption{Topological surface states for the (010) surfaces with (a) Pt and (b) O terminations.
The warmer colors represent the higher surface contribution.
Surface states are labeled by mirror eigenvalues $\pm i$.
For the eigenvalue $+i$ ($-i$), two surface states connect the valence and conduction bands from the right (left) to the left (right).}
\label{Fig_5}
\end{figure}

We performed a slab calculation of $\beta^\prime$-PtO$_2$ using the surface Green's function~\cite{Sancho1985} constructed from the bulk's maximally-localized Wannier functions~\cite{Mostofi2008}.
We stacked unitcells along the $y$ direction and introduced the (010) surface,
whose surface band structure along the $\bar{\mathrm{M}}$\textendash$\bar{\Gamma}$\textendash$\bar{\mathrm{M}}$ direction is shown in Fig.~\ref{Fig_5}.
Depending on the surface terminations, we obtain two distinct dispersions of the surface bands.
For the surface with Pt termination [Fig.~\ref{Fig_5}(a)], the surface states connect the valence band maximum and nearby conduction band minimum.
Whereas for the surface with O termination [Fig.~\ref{Fig_5}(b)], the surface states connect the valence band maximum and conduction band minimum in the other half of the Brillouin zone.
In either case, we observe four surface states crossing the energy gap, which are labeled with $M_{z}$ eigenvalues $\pm i$ in Figs.~\ref{Fig_5}(a) and \ref{Fig_5}(b).
Because there are two surface states for each eigenvalue, we conclude that the  $k_z=0$ plane has a Mirror Chern number $n_M = -2$.
Because our calculation ignores surface potentials and complex surface configurations, the real surface states may show a different dispersion.
Nevertheless, as long as the surface preserves $M_z$, the existence of the surface states, two surface states for each mirror eigenvalue, is guaranteed by topology.
We can compare this result with the surface states of Na$_3$Bi~\cite{Wang:2012ds} and Cd$_3$As$_2$~\cite{Wang:2013is}, which carry nontrivial $\mathbb{Z}_2$ invariants and show odd numbers of crossings in the half of the $k_z=0$ line on the surface Brillouin zone.

\begin{figure}[]
\centering
\includegraphics[width=\columnwidth]{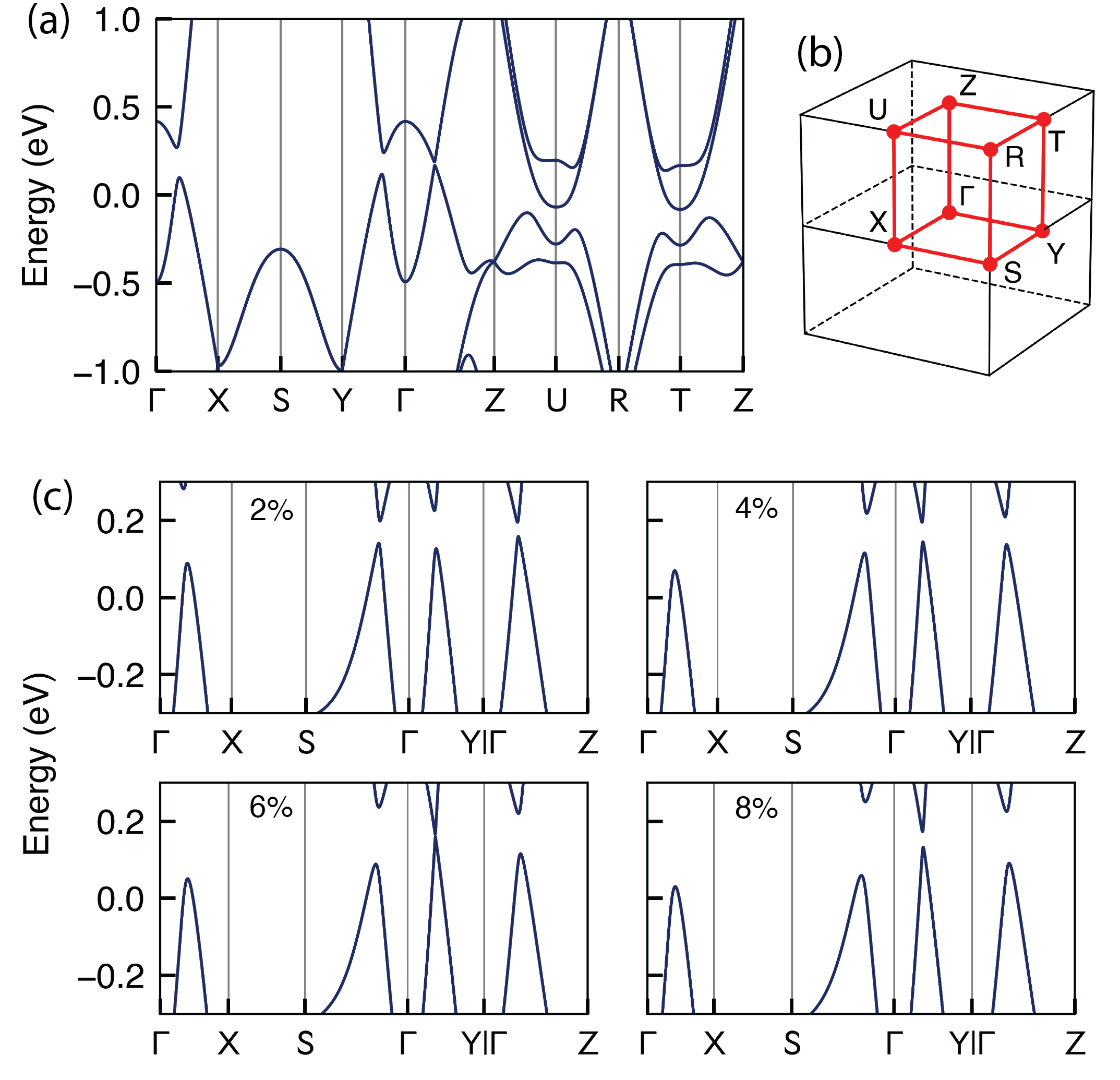}
\caption{(a) Band structure with uniaxial strain. A uniaxial strain of 1~\% along the $x$ direction is applied to $\beta'$-PtO$_2$ of the experimental lattice constants.
(b) Orthorhombic Brillouin zone.
(c) Evolution of the band structure with uniaxial strain.
The band gaps remain open until $\sim 6$~{\%} uniaxial strain.
Then, band gaps close and reopen at the points in the $\Gamma${\textendash}Y line with more strains.}
\label{Fig_6}
\end{figure}

The nontrivial mirror Chern number $n_M = -2$ in the $k_z=0$ plane indicates that $\beta'$-PtO$_2$ can turn into a TCI by some external perturbations.
Because the Dirac points (0,0,$\pm k_d$) are protected by the $\widetilde{C}_{4z}$ symmetry, if we break $\widetilde{C}_{4z}$ while preserving $M_{z}$, the system can become a TCI.
To verify this idea, we apply a uniaxial strain along the $x$ direction to $\beta^\prime$-PtO$_2$ of the experimental lattice constants, and inspect the resulting band structures.
When we apply a uniaxial strain of 1~\%, the Dirac points are immediately gapped as can be seen in Fig.~\ref{Fig_6}(a).
As the other band gaps are not opened or closed, we can identify this phase as a TCI characterized by a mirror Chern number $n_M = -2$.
As we increase the strain, the gapped band structure is preserved until $\sim 6$~\% strain [Fig.~\ref{Fig_6}(c)], which means the system maintains the TCI phase.
With more strains, band gaps close and reopen at the points on the $\Gamma${\textendash}Y line, and the system undergoes a topological phase transition to a normal insulator. 

\subsection{{$\mathbf{k} \cdot \mathbf{p}$} Analysis}

The Dirac nodal lines structure and nontrivial topological invariants of $\beta'$-PtO$_2$ can be explained by constructing a minimal $\mathbf{k} \cdot \mathbf{p}$ Hamiltonian of the system.
First, we consider the case without SOC.
A general two-bands Hamiltonian can be written by a $2 \times 2$ matix $H(\mathbf{k})=\sum_{i=0,x,y,z} a_{i}(\mathbf{k})\tau_i$, where $a_{i}(\mathbf{k})$ is a real function of $\mathbf{k}$, $\tau_0$ is the identity matrix, and $\tau_{x,y,z}$ are the Pauli matrices representing the orbital degree of freedom.
For $\beta'$-PtO$_2$, by analyzing the symmetries of $| \psi_\mathrm{VB} \rangle$ and $| \psi_\mathrm{CB} \rangle$ at the $\Gamma$ point, we can represent the 5 generators of the space group as
$E= \tau_0$,  $C_{2z}=\tau_0$, $\widetilde{C}_{4z}= -\tau_z$, $\widetilde{C}_{2y}= -\tau_0$, and $P= \tau_0$,
in addition we can write $T=K$.
With these symmetries, the Hamiltonian near the $\Gamma $ point is described by the functions
\begin{equation*}
\begin{split}
a_x(\mathbf{k}) &= v_1(k_x^2-k_y^2) \\
a_y(\mathbf{k}) &= 0 \\
a_z(\mathbf{k}) &= v_2-v_3k_z^2-v_4(k_x^2+k_y^2),
\end{split}
\end{equation*}
where $v_i$ is a constant determined by the detailed electronic structures, and we can neglect $a_{0}(\mathbf{k})$ as it does not affect the gap-closing condition.
Because the energy is given by $a_{0}(\mathbf{k}) \pm \sqrt{\sum_{i=x,y,z}a_i(\mathbf{k})^2}$, we solve $a_{x,y,z}(\mathbf{k})=0$ to obtain nodal points.
By letting $k_x = \pm k_y = t/ \sqrt{2}$, we obtain a solution $v_3 k_z^2 + v_4 t^2 = v_2$, which describes ellipses on the (110) and (1$\bar{1}$0) planes consistent with the DFT results in Fig.~\ref{Fig_2}(b).
Including spin degree of freedom, the Hamiltonian is written by a $4\times4$ matrix $H(\mathbf{k})=\sum_{i,j=0,x,y,z} a_{ij}(\mathbf{k})\sigma_i \tau_j$, where $\sigma_0$ is the identity matrix, and $\sigma_{x,y,z}$ are the Pauli matrices for spin.
With SOC, we obtain the same terms as in the spinless case, i.e. $a_{0i}(\mathbf{k})=a_i(\mathbf{k})$, but additionally we have $a_{zy}(\mathbf{k}) = v_5k_xk_y$, which originates from SOC.
Because of this additional term, the ring-shaped nodal lines are gapped except for the two Dirac points at $k_z = \pm \sqrt{v_2/v_3}$.

The nontrivial mirror Chern number in the $k_z=0$ plane can be explained by investigating the $\mathbf{k} \cdot \mathbf{p}$ Hamiltonian around the $(k_0,k_0,0)$ point.
We recall that this point is a Dirac point without SOC, then opens a gap when SOC is introduced.
Now, we restrict the Hamiltonian on the $k_z=0$ plane, and write $H(\mathbf{k})=H(k_x,k_y)$.
To simplify the analysis, we choose $q_x=(k_x+k_y)/\sqrt{2}-\sqrt{2}k_0$ and $q_y=(-k_x+k_y)/\sqrt{2}$, which maps $(k_0,k_0) \to (0,0)$.
In the new coordinate system, the Hamiltonian including SOC is written as
\begin{equation*}
\label{kp}
\begin{split}
a_{0x}(\mathbf{q}) = w_1 q_y,\;
a_{0z}(\mathbf{q}) = w_2 q_x,\;
a_{zy}(\mathbf{q})  =\lambda,
\end{split}
\end{equation*}
where $w_1$, $w_2$, and $\lambda$ are constants~\footnote{
If we include higher-order terms,
we have {$a_{0x}(q_x,q_y) = w_1 q_y +w_3 q_xq_y$} and {$a_{0z}(q_x,q_y) = w_2 q_x+w_4 q_x^2+w_5q_y^2$}.
This is the same form as the type-II semi-Dirac dispersion introduced in Ref. ~\cite{Huang:2015ha}.
We note that the linear term is the most important for the nontrivial Chern number.}.
Because $M_z$ is represented by $-i\sigma_z \tau_0$,
the Hamiltonian is block diagonalized into two eigensectors of $M_z$.
For the $\pm i$ eigensector, the Hamiltonian is written by a $2\times2$ matrix $H_{\pm i}(\mathbf{q})=\mathbf{a}_{\pm i}(\mathbf{q})\cdot \bm{\tau}$, where $\mathbf{a}_{\pm i}(\mathbf{q})=(w_1 q_y, \mp \lambda ,w_2 q_x)$, which describes massive Dirac fermions with a mass term $\lambda$.
The Chern number of the Hamiltonian is explicitly calculated, which gives $-\frac{1}{2} \text{ sgn}\left(w_1 w_2 \lambda\right)$ for the $+i$ eigensector.
Since there are four Dirac points without SOC that are related by the $\widetilde{C}_{4z}$ symmetry, they together give a Chern number $n_{i}=-2 \text{ sgn}\left(w_1 w_2 \lambda\right)$ when SOC opens the gaps.
Whereas for the $-i$ eigensector, the Chern number has the opposite sign because the mass term flips the sign.
Consequently, we classify the $k_z = 0$ plane as a TCI with a mirror Chern number of $\left| 2 \right|$, whose sign should be fixed by the detailed electronic structure.
This nontrivial topological invariant is consistent with the flow of WCCs [Figs.~\ref{Fig_4}(c) and \ref{Fig_4}(d)] and the surface band calculations [Fig.~\ref{Fig_5}].

\section{Discussion and Summary}

We have so far showed that $\beta'$-PtO$_2$ is a topological DSM that possesses a nontrivial mirror Chern number $n_M =-2$.
At this point, it is appropriate to address the shortcomings of the GGA calculations because the correct description of band positions is crucial when Dirac points are created by the band inversion mechanism.
As there is no experimental data regarding the band structure of $\beta^\prime$-PtO$_2$, we have to rely on the first-principles calculations at this moment.
Considering the well-known band gap underestimation of the GGA calculations, we need to verify whether the DSM phase of $\beta'$-PtO$_2$ is retained when we use a higher level exchange-correlation functional.
For this purpose, we employed the HSE functional, where an amount of exact exchange energy, determined by a mixing parameter $\alpha$, is added to the exchange-correlation energy of the GGA.

\begin{figure}[]
\centering
\includegraphics[width=\columnwidth]{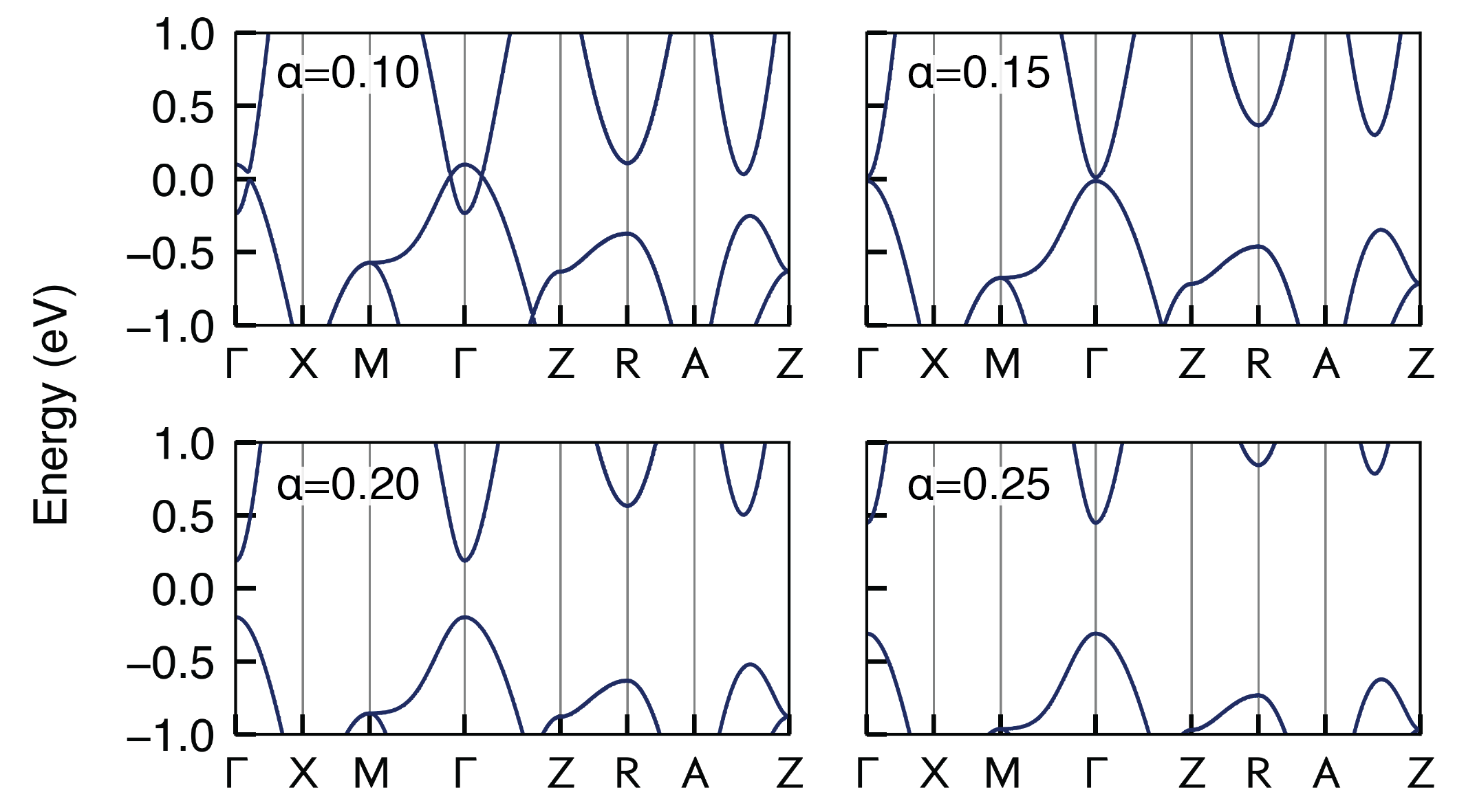}
\caption{HSE band structures with the mixing parameter $\alpha = $ 0.10, 0.15, 0.20, and 0.25.
The band gap increases as $\alpha$ increases.
For $\alpha=0.1 \text{ to } 0.15$ more appropriate for the transition metal systems than $\alpha=0.25$~\cite{Perdew1996,Coulter:2013kv,He2012,Franchini2014}, an ideal DSM phase is observed.}
\label{Fig_7}
\end{figure}

We show the HSE band structures of $\beta^\prime$-PtO$_2$ with the experimental lattice constants in Fig.~\ref{Fig_7}.
Compared with the GGA band structure in Fig.~\ref{Fig_1}(c), the band gap is significantly increased in the HSE calculations.
For $\alpha =$ 0.10, 0.15, 0.20, and 0.25, the band gaps at the $\Gamma$ point are  $-$0.33, 0.025, 0.39, and 0.76 eV, respectively.
Although the $\alpha = 0.25$ is widely used, we emphasize that there is no universal mixing parameter that provides a good description for all materials.
Especially for the rutile and perovskite oxides with transition metal ions, it is known that the smaller mixing parameters $\alpha= 0.1 \text{ to } 0.15$ is more appropriate than the value $\alpha=0.25$~\cite{Perdew1996,Coulter:2013kv,He2012,Franchini2014}.
When the mixing parameter is from 0.1 to 0.15, the band structure of $\beta'$-PtO$_2$ exhibits an ideal 3D DSM phase, where only the Dirac points are pinned at the Fermi level [Fig.~\ref{Fig_7}].
Based on these observations, we hope future experiments will confirm the DSM phase of $\beta^\prime$-PtO$_2$.

\begin{figure}[]
\centering
\includegraphics[width=\columnwidth]{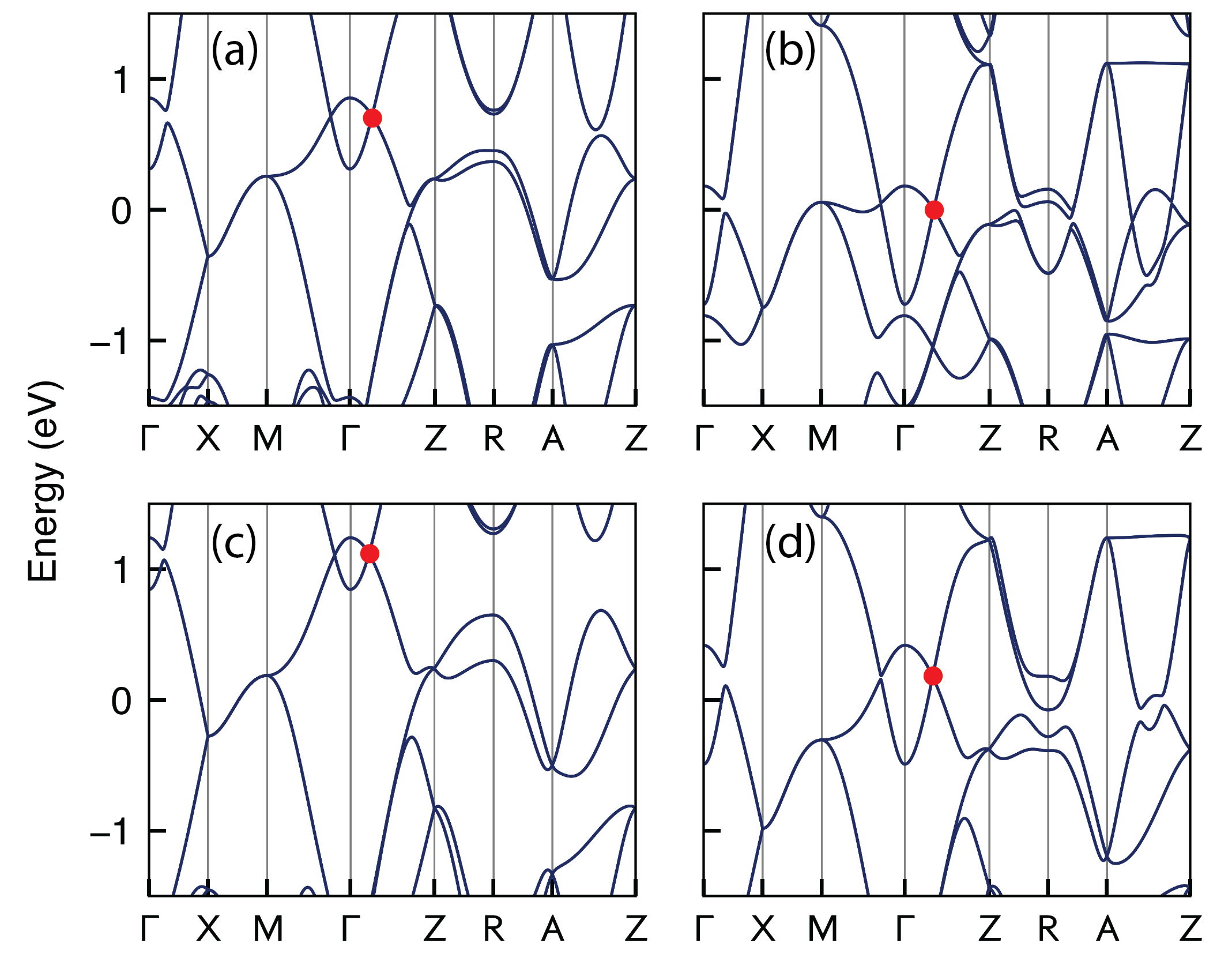}
\caption{GGA + SOC band structures of (a) RhO$_2$ ($a=4.487$~\AA, $c=3.089$~\AA~\cite{Demazeau2006})
(b) PdO$_2$ ($a=4.483$~\AA, $c=3.101$~\AA~\cite{Shaplygin1978}), and 
(c) IrO$_2$ ($a=4.505$~\AA, $c=3.159$~\AA~\cite{Bolzan1997}).
(d) Band structure of $\beta'$-PtO$_2$ ($a=4.485$~{\AA}, $c=3.130$~{\AA}~\cite{Fernandez:1984ga}) is redrawn for comparison.
Red dots denote Dirac points.}
\label{Fig_8}
\end{figure}

Besides $\beta^\prime$-PtO$_2$, possible candidates for 3D topological DSMs include the rutile phase RhO$_2$, PdO$_2$, and IrO$_2$, whose GGA + SOC band structures are shown in Fig.~\ref{Fig_8}.
They share similar band structures, but RhO$_2$ and IrO$_2$ need one electron doping per formula unit because we require 6 electrons in the $d$ orbitals to pin the Dirac points to the Fermi level.
Although such a high level of doping seems impossible, it is recently reported that the rutile structure VO$_2$ can be doped with one electron by using hydrogen atoms~\cite{Yoon:2016bf}, therefore we expect that similar experiments could be possible for RhO$_2$ and IrO$_2$.

In summary, we have theoretically identified the crystalline topological DSM phase in the transition metal rutile oxides as illustrated by $\beta'$-PtO$_2$.
The Dirac points on the $k_z$ axis are protected by the fourfold screw rotation $\tilde{C}_{4z}$ in this class of materials.
Distinct from the known Dirac semimetals such as Cd$_3$As$_2$ and Na$_3$Bi, the topological properties of the system are characterized by the nontrivial mirror Chern number $n_M = -2$ on the $k_z = 0$ plane.
It indicates that a distortion, which breaks the $\tilde{C}_{4z}$ symmetry and preserves the $M_{z}$ symmetry, can make a TCI phase in the system.
Our proposal is likely to lead to a topologically nontrivial DSM phase with strong correlation since the observation is universal among the transition metal rutile oxides with 6 electrons in the $d$ orbitals.

\section*{Acknowledgments}
This work was supported by Institute for Basic Science (IBS) in Korea (Grant No. IBS-R009-D1).
B.-J.Y. was also supported by Basic Science Research Program through the National Research Foundation of Korea (NRF) (Grant No. 0426-20170012, No.0426-20180011), the POSCO Science Fellowship of POSCO TJ Park Foundation (No.0426-20180002), and the U.S. Army Research Office under Grant Number W911NF-18-1-0137.


%

\end{document}